\documentclass[useAMS,usenatbib,usegraphicx]{mn2e}
\def\deg{\ifmmode{^\circ}\else{$^\circ$}\fi}
\def\hGpc{\ifmmode{h^{-1}{\rm Gpc}}\else{$h^{-1}{\rm Gpc}$}\fi}
\def\hkpc{\ifmmode{h^{-1}{\rm kpc}}\else{$h^{-1}{\rm kpc}$}\fi}
\def\hMpc{\ifmmode{h^{-1}{\rm Mpc}}\else{$h^{-1}{\rm Mpc}$}\fi}
\def\hMsun{\ifmmode{h^{-1}M_\odot}\else{$h^{-1}M_\odot$}\fi}

\def\muK{\ifmmode{\mu{\rm{K}}}\else{$\mu$K}\fi}
\def\mum{\ifmmode{\mu{\rm{m}}}\else{$\mu$m}\fi}

\def\simm{\emph{SIM}}
\def\mfrac{$M_{frac} $}

\newcommand{\aj}{{AJ}}
\newcommand{\apj}{{ApJ}}

\newcommand{\apjs}{{ApJS}}
\newcommand{\mnras}{{MNRAS}}

\newcommand{\aap}{A\&A}

\title{The spherical collapse model with shell crossing}

\author[S\'anchez-Conde et al.]
{M.A. S\'anchez-Conde$^{1}$\thanks{E-mail:
masc@iaa.es},
J. Betancort-Rijo$^{2,3}$ \&
F. Prada$^{1}$, 
\\
$^1$
Instituto de Astrofisica de Andalucia (CSIC), E-18008, Granada, Spain
\\
$^2$
Instituto de Astrofisica de Canarias,
C/ Via Lactea s/n, Tenerife, E38200, Spain
\\
$^3$
Facultad de Fisica, Universidad de La Laguna,
Astrofisico Francisco Sanchez, s/n, La Laguna
Tenerife, E38200, Spain
\\
}

\begin{document}

\maketitle

\begin{abstract}
In this work, we study the formation and evolution of dark matter
halos by means of the spherical infall model with shell-crossing. We
present a framework to tackle this effect properly based on the
numerical follow-up, with time, of that individual shell of matter
that contains always the same fraction of mass with respect to the
total mass. In this first step, we do not include angular momentum,
velocity dispersion or triaxiality. Within this framework - named as
the Spherical Shell Tracker (SST) - we investigate the dependence of
the evolution of the halo with virial mass, with the adopted mass
fraction of the shell, and for different cosmologies. We find that our
results are very sensitive to a variation of the halo virial mass or
the mass fraction of the shell that we consider. However, we obtain a
negligible dependence on cosmology. Furthermore, we show that the
effect of shell-crossing plays a crucial role in the way that the halo
reaches the stabilization in radius and the virial equilibrium. We
find that the values currently adopted in the literature for the
actual density contrast at the moment of virialization,
$\delta_{vir}$, may not be accurate enough. In this context, we stress
the problems related to the definition of a virial mass and a virial
radius for the halo. The question of whether the results found here
may be obtained by tracking the shells with an analytic approximation
remains to be explored.
\end{abstract}

\begin{keywords}
{cosmology:theory --- dark matter ---  
large-scale structure of universe --- methods:numerical}
\end{keywords}

\section{Introduction}

In the hierarchical scenario for structure formation in the Universe, the small primordial density fluctuations grow 
due to non-linear gravitational evolution and finally become the first virialized structures (halos). In this 
picture, larger Cold Dark Matter (CDM) halos will be formed by the accretion and merger of those first smaller halos, 
forming in this way massive structures, and so on. This scenario, that constitutes the actual paradigm of 
hierarchical structure formation, is able to explain in general terms the universe that we see today. Yet, we do not 
have a framework or theory capable of reproducing this picture accurately. In this context, N-body cosmological 
simulations are a powerful tool to try to understand the formation and subsequent evolution of CDM halos. They 
constitute a very important help to build any theoretical model and their predictions explain many of different 
observations. 

Basically, there are two analytical approaches that make the problem tractable, although some simplifications have to 
be made and, as it was said, comparison between these analytical studies and simulations are crucial to make 
progress: the Press-Schechter formalism (Press \& Schechter 1974), based on the role of mergers (Nusser \& Sheth 
1999; Manrique et al. 2003), and the spherical infall model (SIM) focused on the understanding of the collapse 
of individual objects. We must note that in the Press \& Schechter formalism, the \emph{SIM} has also been widely 
used, but from a statistical point of view, to treat problems related to mass accretion histories, mass function, 
etc. The \simm, first developed by Gunn \& Gott (1972) and Gunn (1977), describes the collision-less collapse of a 
spherical perturbation in an expanding background. In those two articles, they introduced for the first time the 
cosmological expansion and the role of adiabatic invariance in the formation of individual objects. Fillmore \& 
Goldreich (1984) and Bertschinger (1985) found analytical predictions for the density profiles of collapsed objects 
seeded by scale-free primordial perturbations in a flat universe. Hoffman \& Shaham (1985) generalized these 
solutions to realistic initial profiles in flat and open Friedmann models, and Baarden et al. (1986) (hereafter BBKS) 
improved this work introducing the peak formalism. Later, some studies have been done to include more realistic 
dynamics of the growth process of dark matter halos (e.g. Padmanabhan 1996; Avila-Reese, Firmani \& Hern\'andez 1998; 
Lokas 2000; Subramanian, Cen \& Ostriker 2000). 

In parallel, a large amount of numerical work have been done. Quinn,
Salmon \& Zurek (1986) and Frenk et al.~(1988) obtained isothermal
density profiles ($\rho \propto r^{-2}$) of CDM haloes, and Dubinski
\& Carlberg (1991) and Crone, Evrard \& Richstone (1994) basically
reproduced the predictions of Hoffman \& Shaham (1985) and found some
evidence for no pure power-law density profiles. Later, it was
established that the density profiles of CDM halos have an universal
form (Navarro, Frenk \& White (1996, 1997, hereafter NFW), with $\rho
\propto r^{-1}$ in the inner regions and $\rho \propto r^{-3}$ in the
outskirts, although there is still controversy about the shape of the
profile near the center and recently it has been found that the
profiles flatten out close to $\rho \propto r^{-2}$ beyond $\sim 2$
virial radius. Moore et al.~(1998,1999), amongst others, find $\rho
\propto r^{-1.5}$ in the very center (although in a more recent work
by Graham et al.~(2006) they find $\rho \propto r^{-0.7}$ at 0.1 kpc
from the center) and other authors (Jing \& Suto 2000; Klypin et
al.~2001; Ricotti 2003) find an inner slope ranging from $-1$ to
$-1.5$ depending on halo mass, merger history and
substructure. Concerning the outskirts of dark matter halos, Prada et
al.~(2006) carried out a detailed study and concluded that a 3D
S\'ersic three parameter approximation provides excellent density fits
up to $\sim 2$ times the virial radius, although these profiles differ
considerably from the NFW ones beyond $2$ virial radius.

There are also plenty of works in the literature using the \emph{SIM} to predict the density profiles of dark matter 
halos mainly focused on explaining their central regions. Moreover, the \emph{SIM} has been widely used to obtain 
some quantities specially relevant and directly related to crucial stages in the formation and evolution of CDM 
haloes for different cosmologies, redshifts, etc. A particularly relevant quantity is the value of the overdensity at 
the moment of virialization $\delta_{vir}$ ($\Delta_{vir}$ usually in the literature), where overdensity is defined 
here as a number of times the background density, and its linear counterpart $\delta_{l,vir}$. The values of 
$\delta_{l,vir}$ and $\delta_{vir}$ were obtained introducing the virial theorem into the \emph{SIM} formalism. This 
has important implications in the way we define the virial radius (the radius that attains an overdensity 
$\delta_{vir}$ inside) of dark matter halos in N-body cosmological simulations. $\delta_{vir}$ is conventionally 
chosen to be near $180$ for an Einstein-deSitter cosmology (e.g. Peebles 1980), or $340$ for the 
$\Omega_{\Lambda}=0.7$ cosmology (e.g. Bryan \& Norman 1998; Lokas \& Hoffman 2000).

In the standard derivation of $\delta_{l,vir}$ and $\delta_{vir}$, the typical way to proceed is to assume that a 
shell of matter stabilizes at an epoch twice the time of turn-around (i.e. the time predicted by the standard 
\emph{SIM} to collapse into a point), and in average with a radius that is $1/2$ the turn-around radius (e.g. Peebles 
1980; Lacey \& Cole 1994; Eke, Cole \& Frenk 1996). This $1/2$ factor (in the Einstein-deSitter cosmology; for other 
cosmologies we need to use the Lahav equation, Lahav et al.~1991) is called the \emph{collapse factor}. However, the 
justification to introduce this collapse factor and to suppose the time of virialization as twice the time of 
turn-around, is poor and lack a solid theoretical background. In contrast, in this work we will study the spherical 
collapse without supposing any collapse factor, only taking into account the shell-crossing as the dominant effect. 
The angular momentum and velocity dispersion may also play an important role. The question is: if these effects were 
included in the model, would we obtain the same values for $\delta_{vir}$ and $\delta_{l,vir}$ that those found in 
the most simplistic scenario described by the standard \simm? This issue is one of the aims of the present work.

The main goal of this line of work is to develop a theoretical framework that help us to understand the dynamical 
elements that determine the process of formation of structures (collapsed objects) using spherical symmetry to 
explain main properties of dark matter halos. In this first work we will tackle these questions by means of a "cold" 
collapse, that is, without including the effects of the velocity dispersion and angular momentum. The point is to 
ascertain if the non-uniformity of the density profiles generated via shell-crossing is able to provide the radial 
motions necessary to produce the virialization and stabilization in an appropriate time scale. In a future work, we 
will include the angular momentum and velocity dispersion to go a step further. 

There are some issues that it is worth mentioning and that make this work different from previous works that also 
included the shell-crossing in their formalism (e.g. Lokas \& Hoffman 2000; Nusser 2001; Hiotelis 2002; Ascasibar et 
al.~2004). The way to proceed in these works is to handle the effect of shell-crossing by means of an adiabatic 
invariant, once the standard \emph{SIM} becomes incorrect for late stages of the evolution. This adiabatic invariant, 
also known as \emph{radial action}, makes the problem analytically tractable, and is based on the fact that the 
potential evolves in a time larger than the orbital period of the most inner shells. In contrast, we will study the 
shell-crossing effect doing a follow-up of the radius that contains inside always the same fraction of the virial 
mass. This way to tackle the problem is only one of the possible options, but is essential, for example, in order to 
build and study the relationship between the actual enclosed density contrast $\delta$, defined as $\delta = 
\frac{\rho(<r)-<\rho_m>}{<\rho_m>}$, with $<\rho_m>$ the mean matter density of the Universe, and the linear density 
contrast, $\delta_l$, obtained from the linear theory. Only Gehard Lemson did something similar, although using 
N-body simulations and mainly focused on showing how accurate are the predictions of the standard \emph{SIM} compared 
to his simulations (Lemson 1995). Despite the fact that he showed that the \emph{SIM} is a powerful tool to 
understand the evolution of halos, he never provided detailed quantities and relations for the actual and linear 
density contrasts. The function $\delta_l(\delta)$  is very important to obtain the density profiles of dark matter 
halos, as we discussed in previous works (Prada et al.~2006; Betancort-Rijo et al.~2006). Sheth \& Tormen (2002) 
parametrized this function for the standard \emph{SIM}. The framework presented here will allow us in the near future 
to provide also a simple parametrization for $\delta_l(\delta)$, but taking into account the important effect of 
shell-crossing, together with others relevant effects such as the angular momentum and velocity dispersion. This will 
lead us, for example, to obtain $\delta_{l,vir}$ and its corresponding $\delta_{vir}$, to explain the shape of the 
dark matter density profiles or to shed light on the mass functions. All of this without supposing any collapse 
factor, as pointed before, or other vague assumptions. Nevertheless, it will not be possible to obtain useful 
applications for the moment, since in this first work we will include in our study only the shell-crossing, which is 
the dominant effect. The full treatment will be done and presented in an upcoming work. Here we will provide the 
first results of our theoretical framework related to the role played by the shell-crossing. 

The paper is organized as follows. In section \ref{Sec2} we briefly describe the \emph{SIM} and explain the formalism 
and unities that we will use in the rest of the work. In section \ref{Sec3} we will study in detail the dependence of 
the way that the evolution occurs varying some parameters, in particular, the virial mass of the halo, the fraction 
of mass for a given virial mass, and the cosmology. Section \ref{Sec4} will be specially dedicated to the moments of 
virialization and stabilization according to a given criterion, and their dependence with the same parameters 
described above. We will also emphasize here the difference between these two concepts. Finally, in section 
\ref{Sec5}, we address the main results and ideas of the work, and point the lines for a future work.

\section{The Spherical Shell Tracker Framework}\label{Sec2}
In this section, we first describe the standard \emph{SIM} and its equations, and then we present the formalism that 
we will use in the rest of the work, which will imply to describe the density profile and define our own units. 
This will allow us to handle easily the equations involved. Later, the algorithm that we used to obtain the results 
will be described carefully step by step. All together will be known as the Spherical Shell Tracker Framework 
(\emph{SST}). The main objective of this section is to make easier a possible reproduction and implementation of the 
\emph{SST} framework.

\subsection{The formalism}
In a flat Universe with $\Omega_{\Lambda}=0$, the evolution of a homogeneous spherical (positive) density 
perturbation (the simplest way to tackle the problem of structure formation) with a mass $M$ and radius $R$, is given 
by Newtonian dynamics (as shown by Tolman 1934 and Bondi 1947), provided that $R$ be much smaller than the Hubble 
radius:

\begin{equation}
\frac{d^2R}{dt^2} = \frac{-G~M}{R^2}    \label{eq1}
\end{equation}

Integrating, since $M$ is constant by definition, we obtain:

\begin{equation}
\frac{1}{2}\left(\frac{dR}{dt}\right)^2-\frac{G~M}{R} = E    \label{eq2}
\end{equation}

where $E$ determines whether the sphere expands forever ($E>0$) or it finally contracts ($E<0$).

We can describe in more detail this spherical perturbation with a large number of mass particles, and even it is 
possible and more useful to imagine these particles as concentric shells (thanks to spherical symmetry) that do not 
cross each other, and each of them with a radius $r(j,t)$, where $j$ denotes the shell, which satisfies equation 
(\ref{eq1}):

\begin{equation}
\frac{d^2r(j,t)}{dt^2} = \frac{-G~M(j,t)}{r(j,t)^2}    \label{eq3}   
\end{equation}

where $M(j,t)$ is the enclosed mass for each shell $j$ at time $t$: 

\begin{equation}
M(j,t) = \rho_{crit}\left(\frac{4\pi}{3}r(j,t)^3\right)[1+\delta(r(j,t))]    \label{eq4}
\end{equation}   

being $\rho_{crit}$ the critical density of the Universe, and $\delta(r)$ the actual density contrast within 
$r(j,t)$:

\begin{equation}
\rho_{crit} = \frac{3~H^2}{8\pi~G}; \qquad \delta(r) = \frac{\rho(<r)-<\rho_m>}{<\rho_m>}   \label{eq5}
\end{equation} 

with $H$ the Hubble constant, and $<\rho_m>$ the mean matter density of the Universe.

As long as shell-crossing does not occurs, the actual density contrast is related to the linear one (given by the 
linear theory, see e.g. Padmanabhan 1993) in the Einstein-deSitter cosmology by the formula (Sheth \& Tormen 2002):

\begin{equation}
\delta_{l}(\delta) = \bigg[ 1.68647 - \frac{1.35}{(1+\delta)^{2/3}} - \frac{1.12431}{(1+\delta)^{1/2}} + 
\frac{0.78785}{(1+\delta)^{0.58661}}\bigg]   \label{eq6}
\end{equation}  

The inverse function, $\delta(\delta_l)$, is given by (Patiri et al.~2004):

\begin{displaymath}
\delta(\delta_l)=0.993\big[(1-0.607(\delta _{l}-6.5\times 10^{-3}(1-\theta(\delta_{l})+  
\end{displaymath}
\begin{equation}
+\theta(\delta_l-1.55))\delta_l^2))^{-1.66}-1\big]   \label{eq7}
\end{equation} 
being $\theta$ the step function:
\begin{displaymath}
 \theta(x)=\left \{ \begin{array}{ll}
  1 & \rm{if~x>0}\\
  0 & \rm{if~x\leq 0}
 \end{array} \right.
\end{displaymath}

It is possible to make some simplifications in the equations, choosing in an appropriate manner the value of some 
parameters. In particular, we choose:

\begin{center}
time unit = initial time \\
length unit = initial radius of the protohalo, $R_i$ \\
mass unit = $[1+\delta(R_i)]$
\end{center}

According to these units, and taking into account equations (\ref{eq4}) and (\ref{eq5}), and an Einstein-deSitter 
cosmology (where $H=\frac{2}{3}t^{-1}$), we have:

\begin{equation}
H_i=\frac{2}{3}; \qquad \rho_{crit,i} = \frac{3}{4\pi}; \qquad G=\frac{2}{9}   \label{eq8}
\end{equation}  

where $i$ refers to the initial time.

The Lagrangian radius $q$ for each shell $j$ (i.e. the comoving radius at $t \rightarrow 0$) is related to the 
Eulerian one $r$ by: 

\begin{equation}
q(j)=r_i(j)~[1+\delta(r_i(j))]^{\frac{1}{3}}   \label{eq9}
\end{equation}     

So, for the initial enclosed mass of a shell $j$, we now have simply (in our units and using Eq.(\ref{eq4})):

\begin{equation}
M(j,t_i) = M(j) = q(j)^3   \label{eq10}
\end{equation}  

We must note that this enclosed mass of a shell $j$, $M(j)$, is different from the mass of each shell, 
$M_{shell}(j)$:

\begin{displaymath}
M_{shell}(j)=M(j)-M(j-1)
\end{displaymath}
\begin{equation}
M(j,t)=\sum_{i=1}^n M_{shell}(i) \qquad \textrm{for }  r(i) \leq r(j)   \label{eq10b}
\end{equation}

whit $n$ the total number of shells.

To obtain $q(j)$ using Eq.(\ref{eq9}) we need $\delta^i(r_i(j))$, that
is, the actual density contrast at initial time, and to this end we
need the linear profile at initial time, $\delta_l^i(q(j))$. In this
work we will use the linear profile presented in Prada et al.~(2006)
and Betancort-Rijo et al.~(2006):

\begin{equation}
\delta_l(q)=\delta_{l,vir}\frac{\sigma_{12}(q)}{\sigma(Q)}   \label{eq11} 
\end{equation} 

where $\delta_{l,vir}$ is the linear density contrast at the moment of virialization, $q$ and $Q$ are the Lagrangian 
radii related to $r$ and $R_{vir}$ respectively, that can be obtained using equation (\ref{eq9}), and:

\begin{displaymath}
(\sigma(x))^{2}=\frac{1}{2\pi^{2}}\int_{0}^{\infty}\mid\delta_{k}\mid^{2}W_{T}^2(xk)~k^{2}~dk
\end{displaymath}

\begin{displaymath}
\sigma_{12}=\sigma_{12}(q)=\frac{1}{2\pi^{2}}\int_{0}^{\infty}\mid\delta_{k}\mid^{2}W_{T}(qk)~W_{T}(Qk)~k^2~dk
\end{displaymath}

\begin{equation}
W_{T}(x)=\frac{3(sin~x~-x~cos~x~)}{x^3}   \label{eq13}
\end{equation}  

where $|\delta_{k}|^{2}$ stands for the power spectra of the density fluctuations linearly extrapolated to the 
present. 

There is a good approximation for $\delta_l(q)$:

\begin{displaymath}
  \delta_l(q)=\delta_{l,vir}~exp\left[-b~\bigg(\big(\frac{q}{Q}\big)^2-1\bigg) \right]
\end{displaymath}
\begin{equation}
  b(Q)=-\frac{1}{2}~\frac{d\ln \sigma(x)}{d\ln x} \bigg|_{x=Q}   \label{eq14} 
\end{equation}

with $b$ a constant depending on the mass. In Table \ref{tab1} we present the values of $b$ and $Q$ that we use for 
each mass. Moreover, it is necessary to assign a value to $\delta_{l,vir}$ so that we can use the density profile, 
although one of our final goals is to obtain a precise value for it. In this work we used a $\delta_{l,vir}=1.9$, a 
value which led to good results in previous works (Prada et al.~2006; Betancort-Rijo et al.~2006).

Essentially, the profile given in Eq.(\ref{eq11}) and its approximation in Eq.(\ref{eq14}) takes into account only 
the restriction $\delta_l(Q)=\delta_{l,vir}$. In Hiotelis (2002) and Ascasibar et al.~(2004), a very similar density 
profile was also used, but using the BBKS peak formalism to compute the initial conditions. In a future work we will 
use a more sophisticated density profile that includes also the restriction $\delta_l(q)<\delta_{l,vir}$ for $q>Q$ 
(see Betancort-Rijo et al.~(2006) for a more detailed description), resulting in steeper actual density profiles for 
smaller masses (as confirmed by numerical simulations, see Prada et al.~2006). This fact will probably change 
slightly the results. For the sake of simplicity we preferred to use a simple profile now, although the major 
results of this work will not depend on the assumed profile.

\begin{table} \centering
\caption{Values of $b$ and $Q$ necessary to use the approximation for $\delta_l(q)$ given by (\ref{eq14}).}
\label{tab1} 
\vspace{0.2 cm}
\begin{tabular}{ccc}  

\hline  \hline
M ($\hMsun$) & Q ($\hMpc$) & b   \\
\hline  
\noalign{\smallskip} 
$6.5 \times 10^{10}$ & 0.57 & 0.1889  \\
$5 \times 10^{11}$ & 1.1252 & 0.2202  \\
$3 \times 10^{12}$ & 2.0445 & 0.2544  \\
$2 \times 10^{13}$ & 3.848 & 0.301  \\
$5 \times 10^{14}$ & 11.125 & 0.41  \\
\hline

\end{tabular}
\end{table}

For the Einstein-deSitter cosmology, the $\delta_l^i(q(j))$ profile can be obtained from equation (\ref{eq11}) simply 
rescaling by:

\begin{equation}
  \delta_l^i(q(j)) = \frac{1}{1+z_i}~\delta_l(q(j))  \label{eq15}
\end{equation}

where $z_i$ is the redshift at initial time. We can obtain $\delta^i(r_i(j))$ from $\delta_l^i(q(j))$  using the 
function given in (\ref{eq7}). Inserting this $\delta(\delta_l^i(q(j)))$ in Eq.(\ref{eq9}) we obtain the Lagrangian 
radius for each shell $j$, $q(j)$, and also using Eq.(\ref{eq10}) its enclosed mass $M(j)$.

Once we have the expressions related to the initial conditions and we have presented the density profile, we need the 
equations of evolution. If the shells do not cross each other, then there is an analytical solution for (\ref{eq2}) 
(e.g. Mart\'inez \& Saar 2002) that can be written in the parametric form:

\begin{equation}
r = r_c(1-cos\eta); \quad t = t_c(\eta-sin\eta)   \label{eq16}
\end{equation}  

where:
\begin{equation}
r_c = \frac{GM}{c^2}; \quad t_c = \frac{r_c}{c};  \qquad  \frac{dt}{d\eta}=\frac{R}{c}  \label{eq17}
\end{equation} 

Here $c$ is the velocity of light and there is a change to a non-dimensional variable $\eta$. This solution means that 
the shell expands until it reaches a maximum radius $r_{ta}$, the turn-around radius, at a given time $t_{ta}$, which 
is different for each shell, and after that point the shell starts to contract. We can integrate analytically 
equation (\ref{eq3}) to study the evolution of the spherical density perturbation, at least until the turn-around, 
thanks to the fact that the enclosed mass of the shells do not change with time. Nevertheless, after the turn-around, 
the re-collapse begins and the shell-crossing also starts, so we can not proceed in the same way. At that point, it is 
common to use a prescription based on an adiabatic invariant, to account for this secondary infall and shell-crossing 
(e.g. Lokas \& Hoffman 2000; Nusser 2001; Hiotelis 2002; Ascasibar et al.~2004). On the other hand, one can also 
integrate numerically the equation (\ref{eq3}), computing at each time step the new radius, velocity and enclosed 
mass for each shell. This is the method that we use in our work. Our purpose is to study and to include the 
shell-crossing in our treatment in a natural way, i.e. without making any assumption about the collapse factor, the 
time at which virialization occurs, or any other simplification or approximation. 

We first divide our spherical density perturbation in $n$ equal spherical shells (equivalent to particles), all of 
them with the same thickness, and later we choose the shell $j$ that contain a given fraction of mass of the total 
protohalo. We do so for every time step, from the start of the evolution to the end: we recompute the new enclosed 
mass for each shell at each time step, and we always select that one that contains the fraction of mass we are 
interested in (in that sense, $n$ must be big enough to choose with high precision and without problems at each step 
a shell that contains exactly the required fraction of mass; in our case, $n=3000$ was enough). If we follow for a 
long time the shell related to this fraction of mass, at the end its radius will be almost constant (although the 
corresponding physical shell will change with time), that is, we will reach stabilization (see section 4). Lu et 
al.~(2006) used a similar algorithm, but they divided the halo in equal mass shells, instead of shells with the same 
thickness, as we do. Moreover, their motivations were different, mainly focused on explain the inner shape of the 
density profiles, and they did not carried out a follow-up of any shell in particular.

\subsection{The algorithm}
We now describe the algorithm we used to compute the relevant quantities (radius, velocity and enclosed mass) for 
each shell at each time step.

First, we need to obtain the \emph{initial conditions}:

\begin{enumerate}
\item We divide the protohalo in $n$ equal shells to calculate our array of initial radii, $r_i(j)$, that contains 
the radius $r$ for all the shells. Remember that, in our units, the radius of the total cloud is unity; moreover, $j$ 
increases decreasing the radius, so $r(j=1) = R_i$ (the radius of the whole halo), and $r(j=3000)$ is the radius of 
the deepest shell.
\item In a first approximation, we make $q(j)=r_i(j)$. This will allow us to compute a second and better estimation 
for $q(j)$ using in an appropriate way the relation given by equation (\ref{eq9}), that is:

\begin{equation}
q(j)=r_i(j)[1+\delta(\delta_l^i(q(j)))]^{\frac{1}{3}}   \label{eq18}
\end{equation}  

where we introduce in the right side the $q(j)$ as given by the first approximation. The function $\delta(\delta_l)$ 
is given by Eq.(\ref{eq7}) and $\delta_l^i(q(j))$ is given by (\ref{eq15}).
\item Now, we will use the $q(j)$ obtained in the last step as a new approximation to compute again a better 
estimation 
for $q(j)$, according to eq.(\ref{eq18}).
\item Step (iii) must be repeated until there is no difference between the $q(j)$ that we obtain after each 
iteration, or this difference is less than at least 5\% between two consecutive iterations.
\item With $r_i(j)$ and the last and best estimations for the initial Lagrangian radii of the shells, $q(j)$, we can 
calculate the initial enclosed mass array, $M(j)$, using equation (\ref{eq10}), and the mass of each shell, 
$M_{shell}(j)$, knowing that $M_{shell}(j)=M(j)-M(j-1)$. We will need $M_{shell}(j)$ later to compute the enclosed 
mass array at each time step, since the mass of each shell will be always the same, although the order of the shells 
will be modified.
\item We also need the initial velocity for each shell, $v_i(j)$, as given by the \emph{SIM} (Betancort-Rijo et 
al.~2006):

\begin{eqnarray}
v_i(j) = H_i~r_i(j)~\bigg[1-\frac{1}{3}\frac{1}{1+\delta(\delta_l^i(q(j)))}  \nonumber \\
\times \frac{1}{\frac{d\delta_l(\delta)}{d\delta}|_{\delta=\delta(\delta_l^i(q(j)))}}~\delta_l^i(q(j))\bigg]   
\label{eq19}
\end{eqnarray}   

where $\delta_l(\delta)$ and $\delta(\delta_l)$ are given by Eq.(\ref{eq6}) and Eq.(\ref{eq7}), and 
$\delta_l^i(q(j))$ is given by Eq.(\ref{eq15}). $H_i$ is the Hubble constant at initial time, which in our units is 
$H_i=2/3$.
\item Now it is possible to select the shell that contains the fraction of mass that we are interested in. Our 
studies will be focused on the follow-up of this shell in particular.
\\[0.3cm]
Once we have calculated the initial conditions, we now need to obtain the \emph{equations of evolution}:
\item In our units, equation (\ref{eq3}) can be written as:

\begin{equation}
\frac{d^2r(j,t)}{dt^2} = -\frac{2}{9}\frac{M(j,t)}{r(j,t)^2}   \label{eq20}
\end{equation}

so the equations of the evolution for $r(j)$ and $v(j)$ are:

\begin{equation}
r(j,t+\Delta t) = r(j,t)+v(j,t)~\Delta t    \label{eq21} 
\end{equation}  
\begin{equation}
v(j,t+\Delta t) = v(j,t)-\frac{2}{9}\frac{M(j,t)~\Delta t}{r(j,t)^2}   \label{eq22} 
\end{equation}  

\item We also need to compute how the linear and actual density contrasts evolve with time:

\begin{equation}
\delta_l(j,t)=\delta_l^i(q(j))~t^{\frac{2}{3}}   \label{eq23} 
\end{equation}   
\begin{equation}
\delta(j,t) = \left[[1+\delta(\delta_l^i(q(j)))]\left(\frac{r_i(j)}{r(j)}\right)^3\right]t^2-1    \label{eq24}
\end{equation}  

\item At each time step, we need to recalculate the new enclosed mass array, $M(j)$, using the shell mass array 
$M_{shell}(j)$, since the enclosed mass for a given shell $j$ is equal to:

\begin{equation}
M(j,t)=\sum_{i=1}^n M_{shell}(i) \qquad \textrm{for }  r(i) \leq r(j)   \label{eq25}
\end{equation}

\item At this point, we can select again the shell that contains that
fraction of mass we want to study, to see what happens with its
radius, velocity, and linear and actual density contrasts.
\item For each time step, we will have to repeat (ix) to (xii). 

It is worth mentioning, for possible reproductions of the results,
that we used an optimized temporal step of $0.003$ in our units, which
is good enough to give us robust results of $\delta_l$ and $\delta$
(we checked these values using well known moments of the evolution
like the turn around, where there is no shell-crossing yet). Moreover,
the beginning was set to an initial redshift $z_i=15$, to make sure
that we are still well inside the linear regime, i.e. the initial
value of $\delta_l$ is small enough. However, it must be noted that by
$z=0$ we do not necessarily mean the present time. When we are
considering a given mass scale, $z=0$ corresponds to the time of
virialization of that scale, that is, the time when $\delta_l$ within the
Lagrangian virial radius is $\simeq 1.7$ (more precisely, our $z=0$ is
that one where we have a linear density profile given by
Eq.~\ref{eq14}). So, at z=1/15, $\delta_l$ in that scale is much smaller
than one.  

Furthermore, there is another important question that it is necessary to take into account to implement without 
problems the described algorithm. This is the fact that we have not included yet the effect of the velocity 
dispersion and angular momentum. Therefore, we are in a totally radial (\emph{cold}) collapse and we will have 
problems in the very center of the halo if we simply integrate numerically the equations following this framework. 
When we compute, according to equations (\ref{eq21}) and (\ref{eq22}), the new radius and velocity of a shell which 
is located very near to the center, we can obtain at the following time step a negative radius and a positive 
velocity, which means that actually this shell has crossed through the center and now it goes from the inner regions 
of the halo to the outer ones. In these circumstances, energy is not exactly conserved due to numerical reasons. The 
distance that the shell covers in only one time step is comparable to its radius, which gives a considerable ``leak'' 
of energy. There are different ways to solve this problem; one of them, the solution we chose, is to define a 
parameter $m$ to measure properly this effect and to help us to prevent this lost of energy. If we define for each 
shell the parameter $m$ as $m=\Delta r/r(j,t)$ (where $\Delta r=r(j,t+\Delta t)-r(j,t)$ is the distance that the 
shell has covered along this time step), then there is no problem while $m$ is small enough, but when $m$ reaches a 
larger value, the way to minimize the lost of energy in the process is to change the velocity by its absolute value, 
and keeping intact the value that we have for the radius. Doing so, we \emph{skip} the very center and the loss of 
energy will be minimum. After many attempts, we saw that a value of $m=0.02$ leads to very good results. Once velocity 
dispersion and angular momentum are included, this parameter $m$ will not be necessary.


\end{enumerate}

In the $\Lambda CDM$ cosmology, the formalism and the algorithm are the same, but we must introduce some changes in 
the initial conditions and in the equations of the evolution the spherical to take into account the different 
cosmology with $\Lambda \neq 0$. The equations, modified adequately, are presented in Appendix A.

\section{The evolution of the halo: effect of shell-crossing}   \label{Sec3}

At the beginning of the evolution, we will not obtain any difference using our formalism or using the standard \simm, 
because there is no shell-crossing yet. However, when this effect starts it is clear that this will become false. 
However, the expected deviation with respect to that given by the standard \emph{SIM} may be different if one uses 
different values for the virial mass of the halo, or study different fractions of mass with respect to this virial 
mass, or if we move to a different cosmology. To this end, that is, to quantify in detail how big are the dependences 
on these factors, we carried out a study varying the virial mass of the halo, $M_{vir}$, the fraction of mass related 
to this virial mass, that we call $M_{frac}$, and the cosmology through the value of $\Omega_{\Lambda}$.

\begin{figure}  \centering
\includegraphics[width=8.75cm,height=7.5cm]{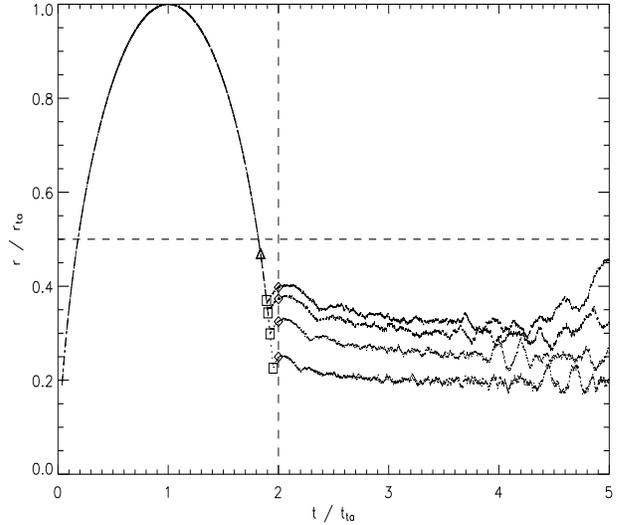}
\caption{Evolution with time of the radius for different $M_{frac}$, for a halo with a virial mass $M_{vir}=3 \times 
10^{12} \hMsun$ and an Einstein-deSitter cosmology. Both radius and time are in units of the turnaround radius and 
time respectively. From down to top, the curves are for $M_{frac}=0.2;~0.5;~0.8;~1.0$. Triangle means $\delta=180$, 
squares the first shell-crossing, and circles indicate the time of collapse according to the standard \simm. The 
horizontal dashed-line corresponds to half the turnaround radius, and the vertical one the time of collapse, i.e. 
twice the turnaround time.}
\label{fig1}
\end{figure}

To illustrate the way in which shell-crossing occurs and affects to the evolution of the halo, in Figure \ref{fig1} 
we show the evolution with time of the radius related to different $M_{frac}$ for the particular case of an 
Einstein-deSitter cosmology and a virial mass of $M_{vir}=3 \times 10^{12} \hMsun$. Both the radius and time are 
expressed in units of the turnaround values (so we can compare between different $M_{frac}$ in the same scale). It is 
worth mentioning that Lemson (1995) presented in the same way data from his simulations and he obtained essentially 
the same results as shown here for the evolution of individual shells.

 Before the first shell-crossing happens, the behaviour of the radii of different $M_{frac}$ is essentially the same. 
This first shell-crossing occurs just before twice the time of turn-around (the time of virialization for the usual 
models), and what we can see is that this first shell-crossing means the beginning of the stabilization in radius, 
which finally occurs some time after that (in next section, we will carry out a detailed study of this process 
together with the virialization). The larger radius oscillations for each curve beyond $\sim 4$ $t/t_{a}$ (see Figure 
1) are only noise due to the growth of numerical errors with time, although in the case of $M_{frac}=1$ the larger 
deviations at larger times are partially and probably due to border effects, i.e. the shell that contains the 
required fraction of mass is near the border of the halo at that time, so there are no enough shells above to obtain 
a good behaviour using our algorithm.

In Table \ref{tab2a} we summarize the results for the linear and
actual density contrasts for a halo with virial mass $M_{vir}=3 \times
10^{12} \hMsun$ and five values of $M_{frac}$ for two different
cosmologies: the Eintein-deSitter and a model with $\Omega_{\Lambda}
\neq 0$. In each case, the corresponding values of $\delta_l$ and
$\delta$ are given for critical or interesting moments of the
evolution, in particular when the first shell-crossing occurs ($FSC$),
when $\delta=180$ ($\Delta$180), $\delta=340$ ($\Delta$340) and when
the collapse occurs ($2TA$) according to the standard \simm, that is,
twice the time of turn-around. The selection of $\Delta$180 and
$\Delta$340 was done because they are the preferred values of
$\delta_{vir}$ in the literature for a flat universe with
$\Omega_{\Lambda}=0$ and $\Omega_{\Lambda}=0.7$ respectively. They
would be also useful to show a possible dependence (or not) of the
function $\delta(\delta_l)$ on the cosmology. We must note that,
although Table \ref{tab2a} is only for a given virial mass, the same
kind of study was done for the evolution of halos with virial masses
$M_{vir}=6.5 \times 10^{10} \hMsun$ and $M_{vir}=5 \times 10^{14}
\hMsun$. We observed the same tendencies in the data and achieved the
same conclusions. In Table \ref{tab3a} we show the values of
$\delta_l$ and $\delta$ that we obtained for the three mentioned
virial masses and for the Eintein-deSitter and the $\Omega_{\Lambda}
\neq 0$ cosmologies, for the particular case in which we fixed \mfrac
=$0.5$.

It is worth mentioning that we did all the calculations to get a
cosmology with $\Omega_{\Lambda}=0.7$ at our nominal $z=0$, so in
general the values of $\delta_l$ and $\delta$ given in the tables are
actually related to other values of the cosmological constant,
i.e. the value of this constant at the corresponding time $FSC$,
$\Delta$180, $\Delta$340 or $2TA$. Because of this fact we give in the
tables the corresponding value of $\beta=\Omega_{m}/\Omega_{\Lambda}$
at the corresponding moment of evolution in the case of a
$\Omega_{\Lambda} \ne 0$ cosmology (in the Einstein-deSitter case it
is not necessary). From the value of $\beta$ we can easily deduce also
the redshift at which that moment of evolution occurs, knowing that
$\beta=\Omega_{m}/\Omega_{\Lambda}=\beta_0~(1+z)^3$, with
$\beta_0=\Omega_{m,0}/\Omega_{\Lambda,0}$ is the value of $\beta$ at
the time corresponding to our $z=0$ (which is $\beta_0=0.429$ for the
$\Omega_m=0.3$, $\Omega_{\Lambda}=0.7$ cosmology). Higher values for
$\beta$ compared to this $\beta_0$ correspond to a virialization or
stabilization that occurs at earlier times ($z>0$), while lower values
represent future times ($z<0$).

\begin{table*} \centering
\caption{Linear ($\delta_l$) and actual ($\delta$) density contrast
values related to some important moments in the evolution of a halo
with a virial mass $M_{vir}=3 \times 10^{12} \hMsun$ and for the
Einstein-deSitter and $\Omega_{\Lambda} \neq 0$
cosmologies. $\beta$ stands for the ratio between $\Omega_m$ and
$\Omega_{\Lambda}$ at the corresponding moment. See text for details,
page 6.}
\label{tab2a} 
\vspace{0.2 cm}
Einstein-deSitter ($\Omega_m=1$, $\Omega_{\Lambda}=0$) \\

\begin{tabular}{ccccccccccccccc}  
\hline \hline
& \multicolumn{2}{c}{$M_{frac}=0.2$} &  & \multicolumn{2}{c}{$M_{frac}=0.5$} & & \multicolumn{2}{c}{$M_{frac}=0.8$} & 
& \multicolumn{2}{c}{$M_{frac}=1.0$} & & \multicolumn{2}{c}{$M_{frac}=1.3$} \\ 
Moment & $\delta_l$ & $\delta$ & & $\delta_l$ & $\delta$ & & $\delta_l$ & $\delta$ & & $\delta_l$ & $\delta$ & & 
$\delta_l$ & $\delta$\\
\hline
\noalign{\smallskip} 
FSC & 1.667 & 1840 & & 1.652 & 770 & & 1.641 & 500 & & 1.634 & 398 & & 1.628 & 329 \\
$\Delta$180 & 1.601 & 180 & & 1.602 & 180 & & 1.602 & 180 & & 1.602 & 180 & & 1.603 & 180 \\
$\Delta$340 & 1.628 & 340 & & 1.628 & 340 & & 1.629 & 340 & & 1.628 & 340 & & 1.628 & 340 \\
2TA & 1.695 & 1426 & & 1.695 & 653 &  & 1.696 & 424 & & 1.696 & 350 & & 1.696 & 289 \\
\hline

\end{tabular}
\end{table*}

\begin{table*} \centering
\label{tab2b} 
\vspace{0.2 cm}
$\Omega_{\Lambda} \neq 0$ \\

\begin{tabular}{cccccccccccccccccccc}  
\hline \hline
& \multicolumn{3}{c}{$M_{frac}=0.2$} &  & \multicolumn{3}{c}{$M_{frac}=0.5$} & & \multicolumn{3}{c}{$M_{frac}=0.8$} & 
& \multicolumn{3}{c}{$M_{frac}=1.0$} & & \multicolumn{3}{c}{$M_{frac}=1.3$} \\ 
Moment & $\delta_l$ & $\delta$ & $\beta$ & & $\delta_l$ & $\delta$ & $\beta$ & & $\delta_l$ & $\delta$ & $\beta$ & & $\delta_l$ & $\delta$ & $\beta$ & & $\delta_l$ & $\delta$ & $\beta$\\
\hline
\noalign{\smallskip} 
FSC & 1.670 & 1728 & 1.686 & & 1.655 & 761 & 1.305 & & 1.644 & 516 & 1.048 & & 1.639 & 427 & 0.910 & & 1.633 & 356 & 0.739 \\
$\Delta$180 & 1.606 & 180 & 1.958 & & 1.606 & 180 & 1.477 & & 1.605 & 180 & 1.162 & & 1.605 & 180 & 1.000 & & 1.605 & 180 & 0.800 \\
$\Delta$340 & 1.632 & 340 & 1.844 & & 1.632 & 340 & 1.382 & & 1.631 & 340 & 1.084 & & 1.631 & 340 & 0.929 & & 1.631 & 340 & 0.741 \\
2TA & 1.698 & 1364 & 1.582 & & 1.698 & 632 & 1.175 & & 1.697 & 446 & 0.912 & & 1.696 & 380 & 0.777 & & 1.696 & 320 & 0.612 \\
\hline

\end{tabular}
\end{table*}

\begin{table*} \centering
\caption{Linear ($\delta_l$) and actual ($\delta$) density contrast values related to some important moments in the 
evolution of a halo for three different virial masses ($M_{vir}=6.5 \times 10^{10} \hMsun$, $M_{vir}=3 \times 10^{12} 
\hMsun$ and $M_{vir}=5 \times 10^{14} \hMsun$) and two different cosmologies (Einstein-deSitter and $\Omega_{\Lambda} \neq 0$). A value of $M_{frac}=0.5$ was set in all the cases. $\beta$ stands for the ratio between $\Omega_m$ and
$\Omega_{\Lambda}$ at the corresponding moment. See text for details, page 6.}
\label{tab3a} 
\vspace{0.2 cm}
Einstein-deSitter ($\Omega_m=1$, $\Omega_{\Lambda}=0$) \\

\begin{tabular}{ccccccccc}  
\hline \hline
& \multicolumn{2}{c}{$M_{vir}=6.5 \times 10^{10} \hMsun$} &  & \multicolumn{2}{c}{$M_{vir}=3 \times 10^{12} \hMsun$} 
& & \multicolumn{2}{c}{$M_{vir}=5 \times 10^{14} \hMsun$} \\ 
Moment & $\delta_l$ & $\delta$ & & $\delta_l$ & $\delta$ & & $\delta_l$ & $\delta$  \\
\hline
\noalign{\smallskip} 
FSC & 1.660 & 1203 & & 1.652 & 770 & & 1.631 & 377   \\
$\Delta$180 & 1.602 & 180 & & 1.602 & 180 & & 1.601 & 180   \\
$\Delta$340 & 1.628 & 340 & & 1.629 & 340 & & 1.628 & 340  \\
2TA & 1.695 & 986 & & 1.695 & 653 &  & 1.694 & 335  \\
\hline

\end{tabular}
\end{table*}

\begin{table*} \centering
\label{tab3b} 
\vspace{0.2 cm}
$\Omega_{\Lambda} \neq 0$ \\

\begin{tabular}{cccccccccccc}  
\hline \hline
& \multicolumn{3}{c}{$M_{vir}=6.5 \times 10^{10} \hMsun$} &  & \multicolumn{3}{c}{$M_{vir}=3 \times 10^{12} \hMsun$} 
& & \multicolumn{3}{c}{$M_{vir}=5 \times 10^{14} \hMsun$} \\ 
Moment & $\delta_l$ & $\delta$ & $\beta$ & & $\delta_l$ & $\delta$ & $\beta$ & & $\delta_l$ & $\delta$ & $\beta$ \\
\hline
\noalign{\smallskip} 
FSC & 1.663 & 1170 & 1.160 & & 1.655 & 761 & 1.305 & & 1.637 & 392 & 1.718  \\
$\Delta$180 & 1.606 & 180 & 1.341 & & 1.606 & 180 & 1.477 & & 1.606 & 180 & 1.848  \\
$\Delta$340 & 1.632 & 340 & 1.254 & & 1.632 & 340 & 1.382 & & 1.632 & 340 & 1.737 \\
2TA & 1.697 & 978 & 1.063 & & 1.698 & 632 & 1.175 & & 1.698 & 338 & 1.489 \\
\hline

\end{tabular}
\end{table*}

Some interesting conclusions can be inferred from the Figure 1 and Tables \ref{tab2a} and \ref{tab3a}. The most 
important ones can be summarized as follows:
\begin{enumerate}
\item $FSC$, the first shell-crossing, occurs earlier as $M_{frac}$ increases.
\item $FSC$ occurs also earlier for larger masses.
\item $FSC$ sets the beginning of the stabilization in radius. 
\item $FSC$ always occurs after $\Delta$180 and $\Delta$340 but always before $2TA$, independently of $M_{vir}$ and \mfrac.
\item $\Delta$180 and $\Delta$340 have essentially the same associated linear density contrasts, no matter the value of 
$M_{frac}$ or $M_{vir}$. This is because in all the cases, there has not been any shell-crossing before $\Delta$180 and 
$\Delta$340, so the standard \emph{SIM} is still valid.
\item Concerning the linear and actual density contrasts for a given $M_{frac}$ and $M_{vir}$, there is no substantial 
difference between the values obtained for different cosmologies.

Specially relevant is the last conclusion, which means that there is no dependence with the cosmology in the values 
of $\delta_l$ and $\delta$, or this dependence is really small and negligible.
\end{enumerate}

Furthermore, we must note here the very high values found for the actual density contrast $\delta$ at the moment of 
the first shell-crossing ($FSC$) and collapse ($2TA$). The reason for that is that there are other important effects, 
together with shell-crossing, involved in the formation and evolution of dark matter halos and that we have not 
included yet in our model. In particular, angular momentum and velocity dispersion will become very relevant and by 
sure will reduce the values that we obtain for $\delta$. In fact, Avila-Reese, Firmani \& Hern\'andez (1998), 
Hiotelis (2002), Ascasibar et al.~(2004) and Shapiro et al.~(2004), amongst others, introduce and study the angular 
momentum and find shallower density profiles in the inner regions, as expected. Hence, it will be absolutely 
necessary to take into account at least these two effects if we want to go a step further in our analysis and if we 
want to obtain a good and accurate parametrization for the function $\delta_l(\delta)$. Nevertheless, the framework 
and algorithm we are using, as well as the conclusions and tendencies we can obtain only including the shell-crossing, 
are totally valid although we can not reach, by now, exact values. Including other effects in our framework, 
specially those ones mentioned above, will be part of a future work.

\begin{figure}  \centering
\includegraphics[width=8.75cm,height=7.5cm]{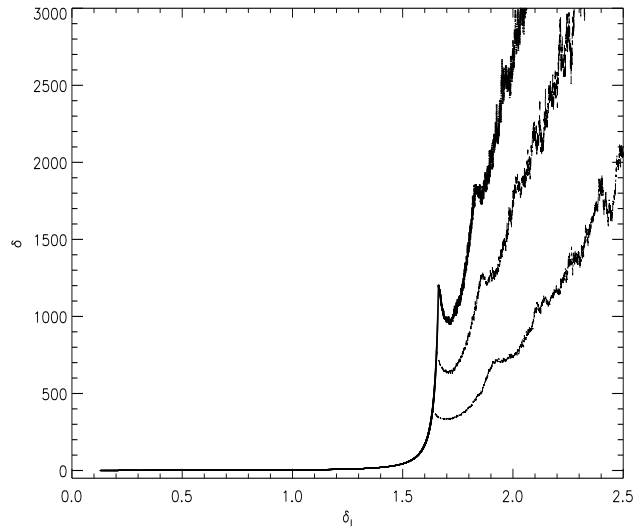}
\caption{The relation $\delta_l - \delta$ for three virial masses. From down to top, the curves correspond to 
$M_{vir}=5 \times 10^{14} \hMsun$, $M_{vir}=3 \times 10^{12} \hMsun$ and $M_{vir}=6.5 \times 10^{10} \hMsun$ (an 
Einstein-deSitter universe and \mfrac=0.5 was used in all the cases).}
\label{fig2}
\end{figure}

\begin{figure}  \centering
\includegraphics[width=8.75cm,height=7.5cm]{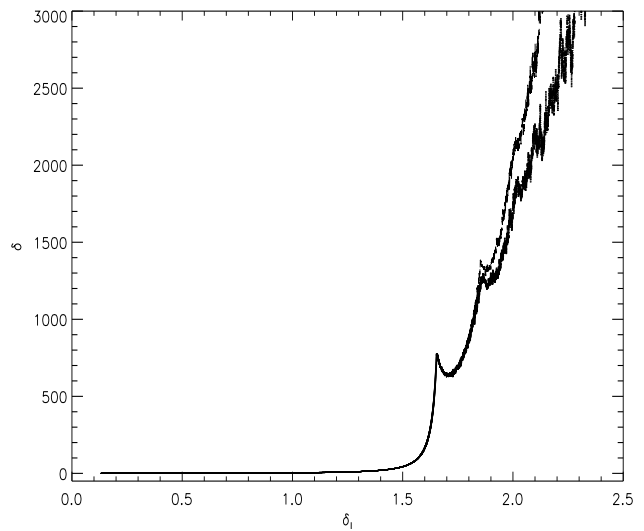}
\caption{The relation $\delta_l - \delta$ for two different cosmologies. From down to top, the curves correspond to 
the Einstein-deSitter case and to the $\Omega_{\Lambda} \neq 0$ cosmology (a virial mass of $M_{vir}=3 
\times 10^{12} \hMsun$ and \mfrac=0.5 was used in all the cases).}
\label{fig3}
\end{figure}

Then, for the moment, we will not be able to provide an exact relation between the linear and actual density 
contrasts, neither a parametric form for the function $\delta_l(\delta)$. However, we can have a look to the relation 
that we obtain at this moment between both density contrasts, and try to extract some conclusions. In Figure 
\ref{fig2}, the function $\delta_l(\delta)$ is represented for the three virial masses under study and for the 
Einstein-deSitter cosmology. Figure \ref{fig3} represents the same function but for different cosmologies, in 
particular for the Einstein-deSitter case and $\Omega_{\Lambda} \neq 0$. The linear region is clearly 
visible in both figures below $\delta_l \sim 1$. In this regime there is no still any difference between the 
different curves and the value of $\delta$ grows very slowly with $\delta_l$, as expected. Then, there is a phase 
where $\delta$ increases very fast for small differences in $\delta_l$, starting from $\delta_l \sim 1.6$ in all the 
cases. From this moment, the dependence with virial mass becomes clearly visible in figure \ref{fig2}, where we 
observe that the smaller the mass, the larger the values of $\delta$ attain for the same value of $\delta_l$. Respect 
to the dependence on different cosmologies, it seems clear (see Figure \ref{fig3}) that this dependence is really 
small, as already mentioned.


\section{Stabilization and Virialization}   \label{Sec4}

In the standard \emph{SIM} and an Einstein-deSitter cosmology, the
value of $\delta_l$ corresponding to the final stage of evolution, to
the so-called \emph{virialization}, is usually taken as
$\delta_{l,vir}=1.686$, that corresponds to an actual density contrast
$\delta_{vir} \approx 180$ (e.g. Peebles 1980). For the
$\Omega_m=0.3$, $\Omega_{\Lambda}=0.7$ cosmology,
$\delta_{l,vir}=1.676$ and $\delta_{vir} \approx 340$ (e.g. Lacey \&
Cole 1994; Eke, Cole \& Frenk 1996; Bryan \& Norman 1998). As pointed
in previous sections, those calculations are based mainly on the
following assumptions:

\begin{enumerate}
\item The halo virializes within a radius that is, on average, a given fraction of its maximum radius (the turnaround 
radius). This fraction is the collapse factor, and is equal to $1/2$ in the Einstein-deSitter cosmology, and in other 
cosmologies it can be inferred from the Lahav equation (Lahav et al.~1991).
\item The time at which virialization occurs is twice the time of turn-around, that is, the time at which the 
collapse happens according to the standard \simm.
\end{enumerate}

Although the values inferred for $\delta_{l,vir}$ and $\delta_{vir}$ in this way are commonly accepted as the correct 
ones and are widely used in the entire literature, the reasons to make the assumptions given above lack a solid 
theoretical base (see section \ref{Sec3}). In fact, there are some works that point to another direction and estimate 
other values of $\delta_{l,vir}$ and $\delta_{vir}$. Jenkins et al.~(2001), for example, find a better agreement with 
the simulations if $\delta_{vir}$ is taken constant for all the cosmologies and near the value that it takes in the 
Einstein deSitter cosmology ($\delta_{vir} \sim 180$). Also Avila-Reese, Firmani \& Hern\'andez (1998) find that a 
different value of $\delta_{l,vir}$ with respect to those obtained using the above assumptions makes better the 
comparison between the analytical Press-Schechter mass distribution and the results of N-body simulations. 

Moreover, there is another important question related to the virialization that should be considered here. In the 
framework of the standard \emph{SIM}, it is possible to apply the virial theorem if we suppose the halo to be an 
isolated system. However, real halos are non-isolated systems, with surrounding material continuously falling or 
escaping from the system. Hence, the virial theorem at least in the standard form could not be applied in this case. 
Despite of this fact, the standard \emph{SIM} together with the virial theorem have been used to obtain the values of 
$\delta_{l,vir}$ and $\delta_{vir}$, and these values have been taken as the references to define the virial radius 
and the virial mass of the halos, which is specially adopted in N-body simulations. Furthermore, this fact has been 
traditionally supported for radial velocity early studies of massive dark matter halos from simulations (Crone et al. 
1994; Cole \& Lacey 1996). These studies apparently showed that the virial radius in this way defined (using 
$\delta_{l,vir}=1.69$ and $\delta_{vir} \sim 180$ in the Einstein-deSitter) constitutes an adequate boundary to 
separate the inner region of the halo in dynamical equilibrium, i.e. that region where the radial velocities are 
zero, from the external region showing infall velocities. The popularization of these ideas came contemporaneously with 
works that defined the virial mass and virial radius in simulations according to these preliminary results (specially since 
the NFW papers). But the fact is that, as recently shown in Prada et al.~(2006), this may not be totally correct. 
Concerning galaxy-size halos, for example, they display all the properties of relaxed objects up to $\sim 3$ 
virial radius and there is no indication of infall of material beyond. Therefore, there is no reason to believe that 
only inside the virial radius, as currently defined, the halo is in equilibrium. In this context, it is important to 
understand the process of virialization more in depth.

In our work, no assumption is done related to the virialization. No collapse factor, no time for virialization 
\emph{a priori} is adopted. Including shell-crossing in the way we do will allow us to obtain $\delta_{l,vir}$ and 
its corresponding $\delta_{vir}$ in a natural way, i.e. studying the evolution of different shells of the halo 
according to the \emph{SST} framework, presented in section 2. It must be noted, however, that these values are still 
preliminary, since it will be necessary to include the effects of angular momentum, velocity dispersion and triaxiality to obtain 
precise and useful values. Nevertheless, this study will be suitable to isolate the role of shell-crossing and will 
be able to extract important conclusions related to the stabilization and virialization. We believe that these 
conclusions will not change when we introduce other physical considerations into the framework.

It is important to note here the difference between these two concepts: stabilization and virialization. The first 
one can be inferred studying the behaviour of the radius of a given shell that contains a given fraction of the 
virial mass with time, as was shown in Figure \ref{fig1}. A criterion must be imposed to say if a given shell reaches 
stabilization or not, and when. The second concept, the virialization, will have to be inferred according to the 
virial theorem. There is no reason why stabilization and virialization should coincide, although instinctively one 
expect that they should be near in time at least.

\subsection{Stabilization} \label{sec41}
In first place, we define the time of stabilization as the time at which the radius of the shell that we are studying 
varies less than a given percent, and during -at least- an interval of time equal to once the time of turnaround. In 
practice, what we do is to choose the moment immediately after the time of first shell-crossing, and we see if there 
is no a variation in radius larger than the maximum variation that we want to impose as our criterion. It must be in 
this way during, at least, a time of turnaround from this moment onward. The value of reference that we take to 
measure the variations in radius is the value of the radius at the initial moment of this interval. If the stated 
percentage of variation of the radius is exceeded in any time within this interval, then we move ahead in time until 
we find an interval of time where the criterion is satisfied. The first moment at which this occurs is our time of 
stabilization, and the value of the radius at the time of stabilization is taken as our radius of stabilization. 

\begin{figure}  \centering
\includegraphics[width=8.75cm,height=7.5cm]{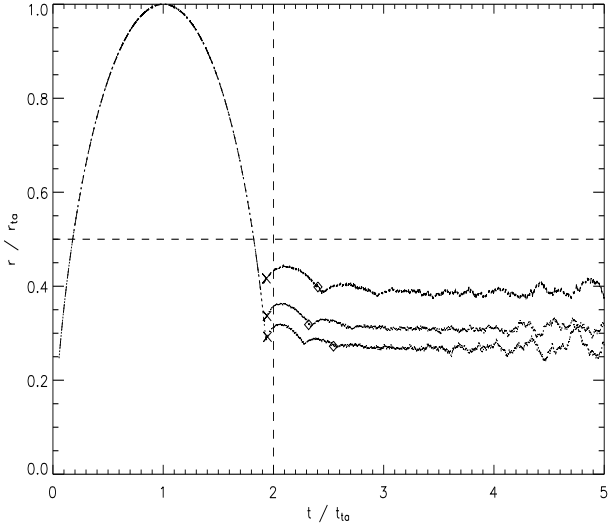}
\caption{Stabilization at 5\% (circles) and 10\% (crosses) for a particular cosmology ($\Omega_m=0.3$, $\Omega_{\Lambda}=0.7$) and 
$M_{frac}=0.5$ for three different virial masses. From down to top, the curves correspond to $M_{vir}=6.5 \times 
10^{10} \hMsun$ $M_{vir}=3 \times 10^{12} \hMsun$ $M_{vir}=5 \times 10^{14} \hMsun$. Again, the horizontal 
dashed-line corresponds to half the turnaround radius, and the vertical one the time of collapse, i.e. twice the 
turnaround time.}
\label{fig4}
\end{figure}

This is illustrated in Figure \ref{fig4}, where the moments of stabilization are shown for a 5\% and 10\% of allowed 
variation of radius, for a particular cosmology ($\Omega_m=0.3$, $\Omega_{\Lambda}=0.7$), a value of $M_{frac}$, and 
for different virial masses. As one can see, in the case of 10\% the stabilization is reached inmediately after the 
first shell-crossing, but the stabilization according to only 5\% of allowed variation of the radius is not reached 
until roughly half of a time unit later (in units of the turnaround time). Moreover, this stabilization remains during more than once the time of turnaround from that moment onward (which is the minimum required by our criterion). Only at late times, where the 
evolution is dominated by numerical noise, the stabilization becomes worse than 5\%. However, the most important 
conclusion is that the stabilization is not reached in any case for a radius that is $1/2$ the radius of turnaround, i.e. the 
value that corresponds to the collapse factor assumed in the standard derivation of $\delta_{l,vir}$ and 
$\delta_{vir}$. This fact constitutes another proof that tell us how inappropriate are the current assumptions done 
about the virialization.

A detailed study was done for different values of $M_{vir}$, $M_{frac}$ and different cosmologies. A summary of this study can 
be found in Appendix B, Tables \ref{tab6a} and \ref{tab7a}. In these tables we present the values for the density 
contrasts that we obtain for two different degrees of stabilization (5\% and 10\%), varying $M_{vir}$, 
$M_{frac}$ and for the Einstein-deSitter and $\Omega_{\Lambda} \neq 0$ cosmologies.

  We must note that the stabilization we have found is numerically
  robust, in the sense that it is independent of the integration
  parameters. In fact, it is in order to be certain that this result
  is not an artifact of approximation used for the dynamics, that we
  have used the direct numerical integration of the equations of
  motion. As we shall show in future work, non-radial orbits and
  triaxiality may modify the form of the stabilization, but here we
  show that shell-crossing itself leads to a stabilization, although
  in the stabilization of actual objects other dynamical factors are
  probably dominant. The origin of the stabilization lies on the fact
  that, as soon as a given shell is crossed, an outward particle flow
  develops which very nearly cancels the initial inward flow (at the
  given shell just before shell crossing). It may also be understood
  as resulting from the radial ``pressure'' which appears after shell
  crossing, that counterbalances gravity. Explaining this fact in the
  simplest possible terms is a rather interesting initiative that we
  shall pursue. However, our intention in this work with respect to
  this point is to establish the fact that cold collapse (i.e. with no
  velocity dispersions) of a spherically symmetric cloud with a
  declining profile does stabilize.

\subsection{Virialization}
Concerning the virialization, what we did was to calculate the kinetic and potential energies related to the shell 
that we want to study. Then, we estimate the degree of agreement with respect to that given by the virial theorem 
(i.e. $U+2K=0$, where $U$ and $K$ are the potential and kinetic energies related to the shell under study) by means 
of the quantity:

\begin{equation}
VIR = \frac{|U+2K|}{2K}  \label{eq26}
\end{equation}

It must be noted that, if the virial theorem was exactly satisfied by the shell at some moment of its evolution, this 
quantity should be at that time equal to zero. 

The algorithm to define the moment and radius of virialization is the same as already described for the stabilization. 
When we find an interval where the degree of virialization that we want to impose is satisfied in every moment inside 
this interval, then we define our time and radius of virialization as those ones corresponding to the beginning of 
the interval considered. Again, as in the case of the stabilization, a detailed study was done for different values 
of $M_{vir}$, $M_{frac}$ and different cosmologies. We summarize the results found in Tables \ref{tab8a} and \ref{tab9a} of 
Appendix B. In these tables we present the values for the density contrasts that we obtain for two different degrees 
of virialization (15\% and 25\%).

There is one issue related to Tables \ref{tab8a} and \ref{tab9a} that it is worth mentioning. In those tables, a 
percent of 15\% and 25\% was set to look for virialization. This was done in this way because we noticed that below 
these percents it is impossible to reach virialization in most cases according to our criterion. A smaller percent 
means a degree of virialization too strong to be satisfied. Nevertheless, it seems that, when the virialization is 
reached using these high percents, the halo has \emph{actually} reached the virialization. This may be inferred from 
the fact that the individual percents that we measure in every moment within the interval considered does not 
decrease monotonically from the beginning of the interval to its end. In fact, what one obtains is a small 
fluctuation around (and near) the high percent that was imposed to find the virialization. That is, the degree of 
virialization does not vary substantially within the whole interval, only small fluctuations are found. There is a 
possible explanation to the fact that we have actually reached virialization but the degree of virialization that we 
find according to our definition is still above 10\% or more. Until now, we have used the standard virial theorem, 
that only involves the kinetic and potential energies. However, because of the fact that we are treating with a 
non-isolated system, with shells of matter continuously going in and going out from the system that we are considering 
as our halo, we should include in the theorem another extra term. This term would be related to the pressures 
involved in the system, and surely may be the explanation and the cause of this ``residual'' percent that we obtain 
in all the cases.

\subsection{Comparison between stabilization and virialization: general considerations}
A summary of our results concerning the degree of both stabilization and virialization for a given shell can be 
found in Tables \ref{tab4a} and \ref{tab5a}. In these tables, we show the degree of stabilization (STA) and 
virialization (VIR) reached for two known moments of evolution, $\Delta$180 and $\Delta$340 (see section 3), for different 
values of $M_{vir}$, $M_{frac}$ and different cosmologies. The degree of virialization, \emph{VIR}, is calculated using Eq.(\ref{eq26}) for moments $\Delta$180 and $\Delta$340. Concerning the degree of stabilization, it was estimated by calculating the following factor:

\begin{equation}
STA = \frac{r_{VIR}-r_{STA}}{r_{STA}}  \label{eq27}
\end{equation}

where $r_{VIR}$ is the value of the radius when $\delta=180$ ($\Delta$180) or when $\delta=340$ ($\Delta$340), and $r_{STA}$ is the 
value of the radius at the moment of stabilization, this one calculated according to the method described in 
section \ref{sec41} and using a value of 10\% for the allowed variation in radius. This factor \emph{STA} can be 
understood as a factor that measures the relative contraction of the radius at times $\Delta$180 or $\Delta$340 with 
respect to that found at the moment of stabilization. 

\begin{table*} \centering
\caption{Values of VIR and STA, calculated according to equations (\ref{eq26}) and (\ref{eq27}), for two moments of evolution $\Delta$180 and $\Delta$340, and for different values of $M_{frac}$ and two cosmologies. A virial mass of $M_{vir}=3 \times 10^{12} \hMsun$ was used in all the cases. $\beta$ stands for the ratio between $\Omega_m$ and $\Omega_{\Lambda}$ at the corresponding moment. See text for details.}
\label{tab4a} 
\vspace{0.2 cm}
Einstein-deSitter ($\Omega_m=1$, $\Omega_{\Lambda}=0$) \\

\begin{tabular}{ccccccccccccccc}  
\hline \hline
& \multicolumn{2}{c}{$M_{frac}=0.2$} &  & \multicolumn{2}{c}{$M_{frac}=0.5$} & & \multicolumn{2}{c}{$M_{frac}=0.8$} & 
& \multicolumn{2}{c}{$M_{frac}=1.0$} & & \multicolumn{2}{c}{$M_{frac}=1.3$} \\ 
Moment & VIR & STA & & VIR & STA & & VIR & STA & & VIR & STA & & VIR & STA \\
\hline
\noalign{\smallskip} 
$\Delta$180 & 6.24 & 0.57 & & 2.96 & 0.43 & & 1.70 & 0.33 & & 1.25 & 0.25 & & 0.83 & 0.19 \\
$\Delta$340 & 5.31 & 0.47 & & 2.52 & 0.31 & & 1.47 & 0.19 & & 1.09 & 0.09 & & 0.31 & 0.09 \\
\hline

\end{tabular}
\end{table*}

\begin{table*} \centering
\label{tab4b} 
\vspace{0.2 cm}
$\Omega_{\Lambda} \neq 0$ \\

\begin{tabular}{cccccccccccccccccccc}  
\hline \hline
& \multicolumn{3}{c}{$M_{frac}=0.2$} & & \multicolumn{3}{c}{$M_{frac}=0.5$} & & \multicolumn{3}{c}{$M_{frac}=0.8$} & 
& \multicolumn{3}{c}{$M_{frac}=1.0$} & & \multicolumn{3}{c}{$M_{frac}=1.3$} \\ 
Moment & VIR & STA & $\beta$ & & VIR & STA & $\beta$ & & VIR & STA & $\beta$ & & VIR & STA & $\beta$ & & VIR & STA & $\beta$ \\
\hline
\noalign{\smallskip} 
$\Delta$180 & 4.23 & 0.49 & 1.959 & & 2.69 & 0.35 & 1.478 & & 1.70 & 0.23 & 1.162 & & 1.43 & 0.19 & 0.999 & & 1.05 & 0.13 & 0.801 \\
$\Delta$340 & 3.88 & 0.39 & 1.843 & & 2.41 & 0.21 & 1.383 & & 1.47 & 0.07 & 1.085 & & 1.29 & 0.05 & 0.930 & & 0.95 & 0.07 & 0.742 \\
\hline

\end{tabular}
\end{table*}

\begin{table*} \centering
\caption{Values of VIR and STA, calculated according to equations (\ref{eq26}) and (\ref{eq27}), for two moments of evolution $\Delta$180 and $\Delta$340, and for three different virial masses and two cosmologies. A value of $M_{frac}=0.5$ was used in all the cases. $\beta$ stands for the ratio between $\Omega_m$ and $\Omega_{\Lambda}$ at the corresponding moment. See text for details.}
\label{tab5a} 
\vspace{0.2 cm}
Einstein-deSitter ($\Omega_m=1$, $\Omega_{\Lambda}=0$) \\

\begin{tabular}{ccccccccc}  
\hline \hline
& \multicolumn{2}{c}{$M_{vir}=6.5 \times 10^{10} \hMsun$} &  & \multicolumn{2}{c}{$M_{vir}=3 \times 10^{12} \hMsun$} 
& & \multicolumn{2}{c}{$M_{vir}=5 \times 10^{14} \hMsun$} \\ 
Moment & VIR & STA & & VIR & STA & & VIR & STA  \\
\hline
\noalign{\smallskip} 
$\Delta$180 & 2.75 & 0.51 & & 2.96 & 0.43 & & 2.00 & 0.23   \\
$\Delta$340 & 2.16 & 0.41 & & 2.52 & 0.31 & & 1.92 & 0.05  \\
\hline

\end{tabular}
\end{table*}

\begin{table*} \centering
\label{tab5b} 
\vspace{0.2 cm}
$\Omega_{\Lambda} \neq 0$ \\

\begin{tabular}{cccccccccccc}  
\hline \hline
& \multicolumn{3}{c}{$M_{vir}=6.5 \times 10^{10} \hMsun$} &  & \multicolumn{3}{c}{$M_{vir}=3 \times 10^{12} \hMsun$} 
& & \multicolumn{3}{c}{$M_{vir}=5 \times 10^{14} \hMsun$} \\ 
Moment & VIR & STA & $\beta$ & & VIR & STA & $\beta$ & & VIR & STA & $\beta$ \\
\hline
\noalign{\smallskip} 
$\Delta$180 & 3.45 & 0.43 & 1.340 & & 2.69 & 0.35 & 1.478 & & 1.16 & 0.17 & 1.849 \\
$\Delta$340 & 2.80 & 0.31 & 1.254 & & 2.41 & 0.21 & 1.383 & & 1.12 & 0.05 & 1.737 \\
\hline

\end{tabular}
\end{table*}

Both tables provide useful results to extract important
conclusions. In first place, one can see that the points $\Delta$180 and
$\Delta$340 are really far from the stabilization and also from the
virialization, although these points are the preferred values for the
moment of virialization in most of the works found in the
literature. In fact, in all the cases the first shell-crossing occurs
even after $\Delta$180 and $\Delta$340, as pointed in section \ref{Sec3}, so it
is not possible that the shell has reached virial equilibrium or
simply stabilization in radius at that moment. Also Lemson (1995)
found similar results using N-body simulations, i.e. the equilibrium
is reached in a longer time respect to that predicted by the standard
\emph{SIM}. In most of the cases, both stabilization and virialization
were obtained far from the value $\delta_l=1.686$ or $\delta_l=1.676$,
the preferred values of $\delta_{l,vir}$ for the Einstein-deSitter and
the $\Omega_m=0.3$, $\Omega_{\Lambda}=0.7$ cosmology respectively (see
Appendix B, Tables \ref{tab6a} to \ref{tab9a}). Concerning the
associated values of the actual density contrast, $\delta$, we find
very high values in most cases. It should be noted, however, that it
is expected that these values decrease substantially when we include
angular momentum and velocity dispersion in the formalism. Therefore,
the values showed in these tables are totally related to the isolated
effect of shell-crossing.

There are also other issues that could be interesting to stress, and that should be explored in more detail in a 
future work:
\begin{enumerate}
\item Concerning $\Delta$180 and $\Delta$340, the degree of both virialization and stabilization are better for larger values 
of $M_{vir}$.
\item Concerning $\Delta$180 and $\Delta$340, the degree of both virialization and stabilization are also better for larger 
values of $M_{frac}$.
\item Concerning $\Delta$180 and $\Delta$340, the degree of both virialization and stabilization are worse in the 
$\Omega_{\Lambda} \neq 0$ cosmology.
\item The moment of stabilization is reached earlier in the $\Omega_{\Lambda} \neq 0$ cosmology.
\item The moment of virialization is reached earlier for smaller values of $M_{vir}$.
\item The moment of virialization is reached earlier for larger values of $M_{frac}$. It is worth mentioning that 
Lemson (1995) found the same from his simulations, i.e. the inner shells reach equilibrium later.
\item The moment of virialization is reached later in the $\Omega_{\Lambda} \neq 0$ cosmology.
\item It seems that there is no coincidence in time between virialization and stabilization, although 
it should substantially depend on the percent that we impose to define both concepts.
\item For all values of $M_{frac}$ and $M_{vir}$, the stabilization in 
radius occurs at a fraction of the turnaround radius that is different from that given by the collapse factor ($1/2$ 
in the Einstein-deSitter case). The same is valid for the $\Omega_{\Lambda} \neq 0$ cosmology.
\end{enumerate}

\section{Summary and future work}   \label{Sec5}

In this work we have studied the effect of shell-crossing in the formation and subsequent evolution of dark matter 
halos. To do that, we have used the spherical collapse model, which has been widely used in the literature for more 
than thirty years to manage and solve questions related to these processes. Despite of the large amount of works that 
have used this model or many others that have improved it by introducing in the formalism more and more complex 
considerations, only a few of them have included the effect of shell-crossing. Moreover, most of these works have 
managed this effect analytically using the adiabatic invariant as a good approximation.

Here we handle the effect of shell-crossing numerically. This allows us to study individually any shell of matter 
involved in the process of formation of the halo. Doing so, we can extract multiple conclusions about the way in 
which this process occurs, like the relation between the linear and actual density contrasts, the process of 
stabilization of a shell of matter, the virialization, etc. Most of these issues have been treated in the present 
work to a greater or lesser extent, although the main goal have been always the developing of an adequate framework 
- named as Spherical Shell Tracker (SST) - in which we can study in depth the shell crossing and also other secondary effects.

It is possible to summarize the main conclusions of this work as follows:

\begin{enumerate}
\item The \emph{SST} framework is adequate to tackle the effect of shell-crossing in a way that allow us to extract 
exact results for different issues related to the evolution of the halo: the way that the radius of a given shell 
evolves with time, the relation between the linear and actual density contrasts, the stabilization, the 
virialization, etc.
\item The shell-crossing by itself is able to produce stabilization and virialization. Nevertheless, for the moment, 
it is not possible to obtain the exact values of the linear and actual density contrasts related to both moments of 
evolution. It is necessary to take into account also other important effects, such as angular momentum, velocity dispersions or triaxiality.
\item Concerning the relation between the linear and actual density contrasts, the dependence of this relation 
with the cosmology is very small and practically negligible. This conclusion is contrary to most of previous works, 
which find in general a large dependence with cosmology. However, the dependence with the virial mass or the fraction 
of virial mass that we consider, is large.  
\item Neither stabilization nor virialization are reached in a time according to that given by the common assumptions 
related to the collapse factor and the time of virialization. In all the cases, we find that both stabilization and 
virialization occur at later times.
\item The values typically used in the literature for $\delta_{l,vir}$ and $\delta_{vir}$ seem to be clearly 
inadequate and incorrect, and are based on not very solid assumptions. In this work, new values of $\delta_{l,vir}$ 
and $\delta_{vir}$ are presented, but only taking into account the effect of shell-crossing. It will be necessary to 
include in our framework other effects also relevant to be able to provide useful and final values for 
$\delta_{l,vir}$ and $\delta_{vir}$.
\end{enumerate}

It is worth to emphasize that this work constitutes only a first step
in our attempt to obtain exact and precise predictions related to the
formation and evolution of dark matter halos. In a future work we plan
to include in the \emph{SST} framework other important effects that it
will be absolutely necessary. In particular, including the angular
momentum and velocity dispersion will be the next step. Furthermore,
in parallel, we will implement a more sophisticated initial density
profile than that used in this work, which fits better that found in
the simulations and could change the results presented here only
slightly. It is also in our mind to use cosmological N-body
simulations, since comparison between both analytical and simulation
studies will be, by sure, crucial to reach a better and deeper
understanding of the processes involved in the formation and evolution
of dark matter halos.

\section*{acknowledgments}

We thank to J.A. Rubi\~no-Mart\'in for useful comments and discussions. M.A.S.C. acknowledges the support of an I3P-CSIC 
fellowship in Granada. M.A.S.C., J.B.R. and F.P. also thank the support of the Spanish AYA2005-07789 grant.

\appendix
\section{The Formalism in the $\Lambda CDM$ cosmology}  \label{apendA}
If we are in a $\Lambda \neq 0$ cosmology, we need to introduce some changes for the initial conditions and in the 
expressions for the evolution of the spherical perturbation, although the formalism and the algorithm are essentially 
the same as presented in section 2, i.e. the \emph{SST} framework. 

The equation for the initial radii of the shells is the same as given by Eq.(\ref{eq18}), but for the velocities the 
correct expression, instead of Eq.(\ref{eq19}), is now:

\begin{displaymath}
v_i(j) = 
\left[1-\frac{1}{3}\frac{1}{1+\delta(\delta_l^i(q(j)))}\frac{a_i~\dot{D}(a)}{D(a)}\bigg|_{a=a_i}\frac{\delta_l^i(q(j)
)}{\frac{d\delta_l(\delta)}{d\delta}\big|_{\delta=\delta(\delta_l^i(q(j)))}}\right]
\end{displaymath}
\begin{equation}
\times ~r_i(j)~\frac{2}{3}~(1+\beta_i)^{\frac{1}{2}}~ln\left[\beta_i^{-\frac{1}{2}}+\sqrt{1+\beta_i^{-1}}\right]  
\label{eqA1}
\end{equation}  

where

\begin{displaymath}
\dot{D}(a) = \frac{dD(a)}{da}
\end{displaymath}
\begin{displaymath}
\beta_i = \frac{\Omega_m}{\Omega_{\Lambda}}~(1+z_i)^3 = \beta_0~(1+z_i)^3
\end{displaymath}

and, 

\begin{displaymath}
a(t) = 
\left[\beta_0^{\frac{1}{2}}\left(\frac{(\beta_i^{-\frac{1}{2}}+\sqrt{1+\beta_i^{-1}})^t-(\beta_i^{-\frac{1}{2}}+\sqrt
{1+\beta_i^{-1}})^{-t}}{2} \right)\right]^{\frac{2}{3}}
\end{displaymath}
\begin{displaymath}
a_i = \frac{1}{1+z_i}
\end{displaymath}
\begin{displaymath}
D(a) = \frac{1}{2~a~f(a)}\int_0^a f^3(a)~da
\end{displaymath}
\begin{equation}
f(a) = \left[1+\Omega_m \left(\frac{1}{a}-1 \right)+\Omega_{\Lambda}(a^2-1)\right]^{-\frac{1}{2}}     \label{eqA2}
\end{equation} 

Note that $\beta_i$ is simply the parameter $\beta_0=\Omega_m / \Omega_{\Lambda}$ but referred to the initial time.

Concerning the initial density profile, now the initial linear density contrast is essentially the same as given 
by Eq.(\ref{eq15}) but now the rescaling factor, $\frac{1}{1+z_i}$, is here replaced by $\frac{D(a_i)}{D(a=1)}$, 
which gives:
\begin{equation}
\delta_l^i(q(j)) = \frac{D(a_i)}{D(a=1)}~\delta_l(q(j))   \label{eqA3}
\end{equation} 

where $\delta_l(q(j))$ is the linear profile given by Eq.(\ref{eq11}).

For the evolution, equations (\ref{eq21}) and (\ref{eq22}) are still valid for the radius and the velocity, but to 
compute the linear and actual density contrast, now we have to include:

\begin{equation}
\delta_l(j,t)=\frac{D(a(t))}{D(a_i)}~\delta_l^i(q(j))  \label{eqA4}
\end{equation} 
\begin{equation}
\delta(j,t) = 
\left[[1+\delta(\delta_l^i(q(j)))]\left(\frac{r_i(j)}{r(j)}\right)^3\left(\frac{a(t)}{a_i}\right)^3\right]-1   
\label{eqA5}
\end{equation}

where $D(a)$ and $a(t)$ are the growing and scale factor respectively, as defined in Eq.(\ref{eqA2}), and $a_i$ 
denotes the scale factor at initial time, given also in Eq.(\ref{eqA2}).

To recompute the enclosed mass at each time step, it is also necessary to take into account the new cosmology, once 
we have calculated $M(j,t)$ in first place according to Eq.(\ref{eq25}), i.e.:

\begin{equation}
M(j,t)(\Lambda \ne 0) = M(j,t) (\Lambda=0) - \frac{2}{\beta_i}r(j,t)^3   \label{eqA6}
\end{equation}

\section{Results obtained for stabilization and virialization}

In Tables \ref{tab6a} to \ref{tab9a} the linear and actual density
contrasts that we obtain concerning the moments of stabilization
and virialization are shown. Tables \ref{tab6a} and \ref{tab7a} refer
to the stabilization whereas Tables \ref{tab8a} and \ref{tab9a} are
related to the virialization. In both cases, the moments of
stabilization (STA) and virialization (VIR) were matched following the
criteria given in Section \ref{Sec4}. In these Tables, linear and
actual density contrast are shown for different values of virial mass,
$M_{vir}$, fraction of virial mass, $M_{frac}$, and for
Einstein-deSitter and $\Omega_{\Lambda} \neq 0$
cosmologies.

\begin{table*} \centering
\caption{Linear and actual density contrasts for two different values
(0.05 and 0.10) of STA, this one defined according to expression
(\ref{eq27}), for different values of $M_{frac}$ and two
cosmologies. A virial mass of $M_{vir}=3 \times 10^{12} \hMsun$ was
used in all the cases. $\beta$ stands for the ratio between $\Omega_m$
and $\Omega_{\Lambda}$ at the corresponding moment. See Section
\ref{Sec4} for details.}
\label{tab6a} 
\vspace{0.2 cm}
Einstein-deSitter ($\Omega_m=1$, $\Omega_{\Lambda}=0$) \\

\begin{tabular}{ccccccccccccccc}  
\hline \hline
& \multicolumn{2}{c}{$M_{frac}=0.2$} &  & \multicolumn{2}{c}{$M_{frac}=0.5$} & & \multicolumn{2}{c}{$M_{frac}=0.8$} & 
& \multicolumn{2}{c}{$M_{frac}=1.0$} & & \multicolumn{2}{c}{$M_{frac}=1.3$} \\ 
STA & $\delta_l$ & $\delta$ & & $\delta_l$ & $\delta$ & & $\delta_l$ & $\delta$ & & $\delta_l$ & $\delta$ & & 
$\delta_l$ & $\delta$\\
\hline
\noalign{\smallskip} 
0.05 & - & - & & - & - & & - & - & & - & - & & - & - \\
0.10 & 1.98 & 3993 & & 2.00 & 1789 & & 2.05 & 1223 & & 1.89 & 700 & & 1.90 & 537 \\
\hline

\end{tabular}
\end{table*}

\begin{table*} \centering
\label{tab6b} 
\vspace{0.2 cm}
$\Omega_{\Lambda} \neq 0$ \\

\begin{tabular}{cccccccccccccccccccc}  
\hline \hline
& \multicolumn{3}{c}{$M_{frac}=0.2$} &  & \multicolumn{3}{c}{$M_{frac}=0.5$} & & \multicolumn{3}{c}{$M_{frac}=0.8$} & 
& \multicolumn{3}{c}{$M_{frac}=1.0$} & & \multicolumn{3}{c}{$M_{frac}=1.3$} \\ 
STA & $\delta_l$ & $\delta$ & $\beta$ & & $\delta_l$ & $\delta$ & $\beta$ & & $\delta_l$ & $\delta$ & $\beta$ & & $\delta_l$ & $\delta$ & $\beta$ & & $\delta_l$ & $\delta$ & $\beta$ \\
\hline
\noalign{\smallskip} 
0.05 & 1.91 & 3306 & 0.959 & & 1.85 & 1298 & 0.813 & & 1.86 & 897 & 0.577 & & 1.86 & 741 & 0.483 & & 1.86 & 621 & 0.362 \\
0.10 & 1.67 & 1496 & 1.697 & & 1.67 & 703 & 1.269 & & 1.67 & 463 & 0.977 & & 1.67 & 390 & 0.829 & & 1.67 & 324 & 0.663 \\
\hline

\end{tabular}
\end{table*}

\begin{table*} \centering
\caption{Linear and actual density contrasts for two different values
  (0.05 and 0.10) of STA, this one defined according to expression
  (\ref{eq27}), for three different virial masses and two
  cosmologies. A value of $M_{frac}=0.5$ was used in all the
  cases. $\beta$ stands for the ratio between $\Omega_m$ and
  $\Omega_{\Lambda}$ at the corresponding moment. See Section
  \ref{Sec4} for details.}
\label{tab7a} 
\vspace{0.2 cm}
Einstein-deSitter ($\Omega_m=1$, $\Omega_{\Lambda}=0$) \\

\begin{tabular}{ccccccccc}  
\hline \hline
& \multicolumn{2}{c}{$M_{vir}=6.5 \times 10^{10} \hMsun$} &  & \multicolumn{2}{c}{$M_{vir}=3 \times 10^{12} \hMsun$} 
& & \multicolumn{2}{c}{$M_{vir}=5 \times 10^{14} \hMsun$} \\ 
STA & $\delta_l$ & $\delta$ & & $\delta_l$ & $\delta$ & & $\delta_l$ & $\delta$  \\
\hline
\noalign{\smallskip} 
0.05 & - & - & & - & - & & - & -   \\
0.10 & - & - & & 2.00 & 1789 & & 1.87 & 610  \\
\hline

\end{tabular}
\end{table*}

\begin{table*} \centering
\label{tab7b} 
\vspace{0.2 cm}
$\Omega_{\Lambda} \neq 0$ \\

\begin{tabular}{cccccccccccc}  
\hline \hline
& \multicolumn{3}{c}{$M_{vir}=6.5 \times 10^{10} \hMsun$} &  & \multicolumn{3}{c}{$M_{vir}=3 \times 10^{12} \hMsun$} 
& & \multicolumn{3}{c}{$M_{vir}=5 \times 10^{14} \hMsun$} \\ 
STA & $\delta_l$ & $\delta$ & $\beta$ & & $\delta_l$ & $\delta$ & $\beta$ & & $\delta_l$ & $\delta$ & $\beta$ \\
\hline
\noalign{\smallskip} 
0.05 & 1.93 & 2727 & 0.569 & & 1.85 & 1298 & 0.813 & & 1.89 & 680 & 0.944  \\
0.10 & 1.67 & 1105 & 1.137 & & 1.67 & 703 & 1.269 & & 1.67 & 352 & 1.602 \\
\hline

\end{tabular}
\end{table*}

\begin{table*} \centering
\caption{Linear and actual density contrasts for two different values
  (0.15 and 0.25) of VIR, this one defined according to expression
  (\ref{eq26}), for different values of $M_{frac}$ and two
  cosmologies. A virial mass of $M_{vir}=3 \times 10^{12} \hMsun$ was
  used in all the cases. $\beta$ stands for the ratio between
  $\Omega_m$ and $\Omega_{\Lambda}$ at the corresponding moment. See
  Section \ref{Sec4} for details.}
\label{tab8a} 
\vspace{0.2 cm}
Einstein-deSitter ($\Omega_m=1$, $\Omega_{\Lambda}=0$) \\

\begin{tabular}{ccccccccccccccc}  
\hline \hline
& \multicolumn{2}{c}{$M_{frac}=0.2$} &  & \multicolumn{2}{c}{$M_{frac}=0.5$} & & \multicolumn{2}{c}{$M_{frac}=0.8$} & 
& \multicolumn{2}{c}{$M_{frac}=1.0$} & & \multicolumn{2}{c}{$M_{frac}=1.3$} \\ 
VIR & $\delta_l$ & $\delta$ & & $\delta_l$ & $\delta$ & & $\delta_l$ & $\delta$ & & $\delta_l$ & $\delta$ & & 
$\delta_l$ & $\delta$ \\
\hline
\noalign{\smallskip} 
0.15 & 2.37 & 8344 & & 2.20 & 2513 & & 2.08 & 1224 & & 2.01 & 784 & & 1.91 & 551 \\
0.25 & 2.25 & 6843 & & 2.09 & 2080 & & 1.97 & 903 & & 1.90 & 731 & & 1.81 & 347 \\
\hline

\end{tabular}
\end{table*}

\begin{table*} \centering
\label{tab8b} 
\vspace{0.2 cm}
$\Omega_{\Lambda} \neq 0$ \\

\begin{tabular}{cccccccccccccccccccc}  
\hline \hline
& \multicolumn{3}{c}{$M_{frac}=0.2$} &  & \multicolumn{3}{c}{$M_{frac}=0.5$} & & \multicolumn{3}{c}{$M_{frac}=0.8$} & 
& \multicolumn{3}{c}{$M_{frac}=1.0$} & & \multicolumn{3}{c}{$M_{frac}=1.3$} \\
VIR & $\delta_l$ & $\delta$ & $\beta$ & & $\delta_l$ & $\delta$ & $\beta$ & & $\delta_l$ & $\delta$ & $\beta$ & & $\delta_l$ & $\delta$ & $\beta$ & & $\delta_l$ & $\delta$ & $\beta$ \\
\hline
\noalign{\smallskip} 
0.15 & - & - & - & & - & - & - & & - & - & - & & - & - & - & & - & - & - \\
0.25 & 2.65 & 23382 & 0.119 & & 2.47 & 8359 & 0.119 & & 2.33 & 4939 & 0.119 & & 2.25 & 3119 & 0.119 & & 2.14 & 2027 & 0.119 \\
\hline

\end{tabular}
\end{table*}

\begin{table*} \centering
\caption{Linear and actual density contrasts for two different values
  (0.15 and 0.25) of VIR, this one defined according to expression
  (\ref{eq26}), for three different virial masses and two
  cosmologies. A value of $M_{frac}=0.5$ was used in all the
  cases. $\beta$ stands for the ratio between $\Omega_m$ and
  $\Omega_{\Lambda}$ at the corresponding moment. See text for
  details.}
\label{tab9a} 
\vspace{0.2 cm}
Einstein-deSitter ($\Omega_m=1$, $\Omega_{\Lambda}=0$) \\

\begin{tabular}{ccccccccc}  
\hline \hline
& \multicolumn{2}{c}{$M_{vir}=6.5 \times 10^{10} \hMsun$} &  & \multicolumn{2}{c}{$M_{vir}=3 \times 10^{12} \hMsun$} 
& & \multicolumn{2}{c}{$M_{vir}=5 \times 10^{14} \hMsun$} \\ 
VIR & $\delta_l$ & $\delta$ & & $\delta_l$ & $\delta$ & & $\delta_l$ & $\delta$ \\
\hline
\noalign{\smallskip} 
0.15 & - & - & & 2.20 & 2513 & & 2.91 & 3886  \\
0.25 & 1.92 & 2132 & & 2.11 & 2132 & & 2.72 & 2378  \\
\hline
\end{tabular}
\end{table*}

\begin{table*} \centering
\label{tab9b} 
\vspace{0.2 cm}
$\Omega_{\Lambda} \neq 0$ \\

\begin{tabular}{cccccccccccc}  
\hline \hline
& \multicolumn{3}{c}{$M_{vir}=6.5 \times 10^{10} \hMsun$} &  & \multicolumn{3}{c}{$M_{vir}=3 \times 10^{12} \hMsun$} 
& & \multicolumn{3}{c}{$M_{vir}=5 \times 10^{14} \hMsun$} \\ 
VIR & $\delta_l$ & $\delta$ & $\beta$ & & $\delta_l$ & $\delta$ & $\beta$ & & $\delta_l$ & $\delta$ & $\beta$ \\
\hline
\noalign{\smallskip} 
0.15 & 2.19 & 6056 & 0.268 & & - & - & - & & 2.53 & 3931 & 0.161  \\
0.25 & 2.08 & 4141 & 0.376 & & 2.47 & 8359 & 0.119 & & 2.42 & 3177 & 0.237  \\
\hline

\end{tabular}
\end{table*}

\end{document}